\documentclass[twocolumn,showpacs,superscriptaddress,prl,aps,floatfix]{revtex4-1}

\usepackage{graphicx}
\usepackage{rotating}
\usepackage{amsmath}
\usepackage[usenames,dvipsnames]{color}
\usepackage{float}
\usepackage{rotating}
\usepackage{upgreek}

\usepackage[colorlinks,linkcolor=blue,citecolor=green]{hyperref}
\usepackage[T1]{fontenc}
\usepackage{bm}

\makeatletter

\newcommand{\Rmnum}[1]{\expandafter\@slowromancap\romannumeral #1@}

\newcommand{\DC}[1]{\textcolor{red}{To Domenico and Cesare:}}

\makeatletter

\hfuzz=\maxdimen
\usepackage[draft]{changes} 
\usepackage[normalem]{ulem}
\definechangesauthor[name={Jiangang}, color=red]{JH}

\begin{document}

\title{Assessing the performance of self-consistent hybrid functional for band gap calculation in oxide semiconductors}

\author{Jiangang He}
\affiliation{University of Vienna, Faculty of Physics and Center for Computational Materials Science, Vienna, Austria}
\affiliation{Department of Materials Science and Engineering, Northwestern University, Evanston, IL USA}

\author{Cesare Franchini}
\email{cesare.franchini@univie.ac.at}
\affiliation{University of Vienna, Faculty of Physics and Center for Computational Materials Science, Vienna, Austria}

\date{\today}


\begin{abstract}
In this paper we assess the predictive power of the self-consistent hybrid functional scPBE0 in calculating the band gap of oxide semiconductors.
The computational procedure is based on the self-consistent evaluation of the mixing parameter $\alpha$ by means of an iterative calculation of the static dielectric constant using the perturbation expansion after discretization (PEAD) method and making use of the relation 
$\alpha = 1/\epsilon_{\infty}$.
Our materials dataset is formed by 30 compounds covering a wide range of band gaps and dielectric properties, and includes materials with a 
wide spectrum of application as thermoelectrics, photocatalysis, photovoltaics,  transparent conducting oxides, and refractory materials. 
Our results show that the scPBE0 functional provides  better band gaps than the non self-consistent hybrids PBE0 and HSE06, but scPBE0 does not show significant improvement on the description of the static dielectric constants. Overall, the scPBE0 data exhibit a mean absolute percentage error  
of 14 \% (band gaps) and 10 \% ($\epsilon_\infty$).  
For materials with weak dielectric screening and large excitonic biding energies scPBE0, unlike PBE0 and HSE06, overestimates the band gaps, but 
the value of the gap become very close to the experimental value when excitonic effects are included (e.g. for SiO$_2$).
However, special caution must be given to the compounds with small band gaps due to the tendency of scPBE0 to overestimate the dielectric constant
in proximity of the metallic limit.
\end{abstract}

\maketitle

\section{Introduction}
Oxides are one of the most physical complicated and technological important systems, exhibiting a variety of structures and interesting properties.
Metal oxides have broaden applications in many relevant areas including superconductivity~\cite{emery1987theory}, ferroelectricity~\cite{cohen1992origin}, dielectric~\cite{shannon1993dielectric}, transparent conducting oxides~\cite{ginley2000transparent},
thermoelectricity~\cite{fergus2012oxide}, photocatalysis~\cite{linsebigler1995photocatalysis}, and photovoltaic materials~\cite{fortunato2007transparent}.
One of the fundamental properties of materials, which is functional for virtually any type of application is the band gaps.
The theoretical understanding of oxides and the accurate estimation of the band gap is however a very difficult task. This is due to the nature of the metal-oxygen bonding that can very between nearly ionic to highly covalent or metallic~\cite{rao1989transition}, and by the generally high degree of electronic correlation, which poses serious problems to conventional computational methods.

With the rapid development of \emph{first principles} computational approaches to solve the many-body Schr\"odinger equation for extended systems, in particular the density functional theory (DFT), many materials-specific properties have been calculated without relying on any empirical parameter~\cite{Lejaeghereaad3000}. However, it is still very difficult to calculate band gap accurately within the DFT framework using standard semilocal exchange-correlation (XC) functionals such as the local density approximation (LDA)~\cite{PhysRevLett.45.566} and the generalized gradient approximation (GGA)~\cite{PhysRevB.46.6671}. This is a particularly pressing problem, considering the fundamental relevance and the technological importance of bands gaps. Both LDA and GGA inevitably underestimate the value of band gaps, leading to values which are significantly smaller than the measured ones. The reason for this failure lays on the approximate form of the XC functionals, ultimately related to the lack of
derivative discontinuities of the XC energy when the electron number increases through an integer point~\cite{PhysRevLett.51.1884,PhysRevLett.51.1888}.
In fact, in DFT the value of the band gap is obtained from the one-particle eigenvalues and the finite correction arising from this derivative discontinuity is not taken into account. Many different approaches have been proposed to reducing this intrinsic inaccuracy of DFT functionals: the weighted density approximation~\cite{wda}, the DFT+U approximation~\cite{dftu}, the self-interaction correction method~\cite{sic},
the screened exchange approximation~\cite{sx}, the optimized effective potential~\cite{oep}, generalized Kohn-Sham schemes~\cite{gks} (GKS), various meta-GGA potentials~\cite{metagga}, and hybrid functionals~\cite{hfs}, to name the most representative ones~\cite{RevModPhys.80.3}. Alternative and more advanced ways to go beyond DFT are the GW approximation~\cite{gw}, in which the self energy of a many-body system of electrons is computed explicitly by making use of the single particle Green's function G and the screened Coulomb interaction W. 

Similarly to DFT, also the Hartree-Fock theory (HF) does not provide a good prediction of band gap, but in this case this failure arises from the lack of proper correlation (C) effects in the single Slater determinant picture, which leads to much large values of the band gap. In the HF framework, this limitation is cured by incorporating electronic correlation as done in the so-called post-HF methods (M{\o}ller-Plesset perturbation theory~\cite{mp}, Configuration interaction~\cite{ci}, and Coupled cluster~\cite{cc}).

In this variety of methods, hybrid functionals turned to be a valid compromise between computational cost and accuracy and have been increasing used in the computational materials science community~\cite{PhysRevB.86.235117,0953-8984-26-25-253202,0953-8984-20-6-064201,PhysRevB.89.195112,PSSB:PSSB201046195, PhysRevB.85.115129,PSSB:PSSB201046303,PhysRevB.84.115114,PhysRevB.85.195135,PhysRevB.91.155201,doi:10.1021/acs.jpclett.6b01807,doi:10.1021/acs.jpclett.6b01807}.

Hybrid functionals are typically seen as a suitable mixing between HF and DFT, obtained by replacing a portion of the exchange functional
with the exact HF exchange (X). 
The fundamental aspect at the core of  the construction of a hybrid functional is the adiabatic connection formula, which continuously transforms the non-interacting particle system to the physical interacting one:

\begin{equation}
{\rm E_{ XC}=\int_0^1 E_{XC,\lambda}d\lambda} 
 \label{eq:ad}
\end{equation}

where E$_{\rm XC}$ is the Kohn-Sham (KS) exchange-correlation energy and $\lambda$ is the coupling parameter that tunes the electron Coulomb potential from the KS ($\lambda=0$)  to the fully interacting ($\lambda=1$) limit. The first hybrid functional originally proposed by Becke in 1993~\cite{doi:10.1063/1.464913} is based on an approximate evaluation of the coupling-constant integral (Eq.\ref{eq:ad}) using a simple 
linear interpolation:

\begin{equation}
 {\rm E_{ XC}=\int_0^1 E_{XC,\lambda}d\lambda~\rightarrow~E_{XC}^{Hybrid}=\frac{1}{2}E_X^{HF}+\frac{1}{2}E_{XC}^{LDA}}
 \label{eq:ad2}
\end{equation}

$E_{XC}^{Hybrid}$ is generally referred to as the half-half hybrid.
After proposing the half-half hybrid, Becke introduced a parametric hybrid functional including exact exchange and local (LDA) and
gradient-corrected (GGA) exchange and correlation terms that has become very popular in the quantum chemistry community with the abbreviation
B3LYP~\cite{Becke1993b,Stephens}. The B3LYP depends on three parameters and incorporates only 20\% of the exact HF exchange and
have the following form:

\begin{widetext}
\begin{equation*}
\label{b3lyp}
{\rm E_{ XC}^{\rm B3LYP} = E_{XC}^{LDA} + \alpha_1(E_X^{HF} - E_{X}^{LDA}) 
+ \alpha_2(E_{X}^{GGA} - E_{X}^{LDA}) + \alpha_3(E_{C}^{GGA} - E_{C}^{LDA})}
\end{equation*}
\end{widetext}

where the three mixing parameters $\alpha_1=0.2$, $\alpha_2=0.72$, and
$\alpha_3=0.81$ are determined by fitting experimental atomization energies,
electron and proton affinities and ionization potentials of the molecules
in Pople's G1 data set. B3LYP has been intensively and successfully adopted for atomic and molecular calculations, but
its application to periodic systems is not equally satisfactory, because B3LYP functional does not reproduce the correct exchange
correlation energy for the free-electron gas. This is particularly problematic for metals
and heavier elements, beyond the 3$d$ transition metal series~\cite{Paier2007}.

A more appropriate hybrid functional for solid-state applications is the PBEh proposed by Perdew, Burke, and Ernzerhof~\cite{pbe0a}
(also referred to as PBE0)~\cite{pbe0b}, which reproduces the homogeneous electron gas limit and significantly outperforms B3LYP in solids,
especially in the case of systems with itinerant character (metals and small gap semiconductors)~\cite{Paier2007}:

\begin{equation}\label{pbe0}
{\rm E_{XC}^{\rm PBE0} = E_{XC}^{PBE} + \alpha(E_X^{HF} + E_X^{PBE})}
\end{equation}

\begin{figure}
        \includegraphics[clip,width=1.0\linewidth]{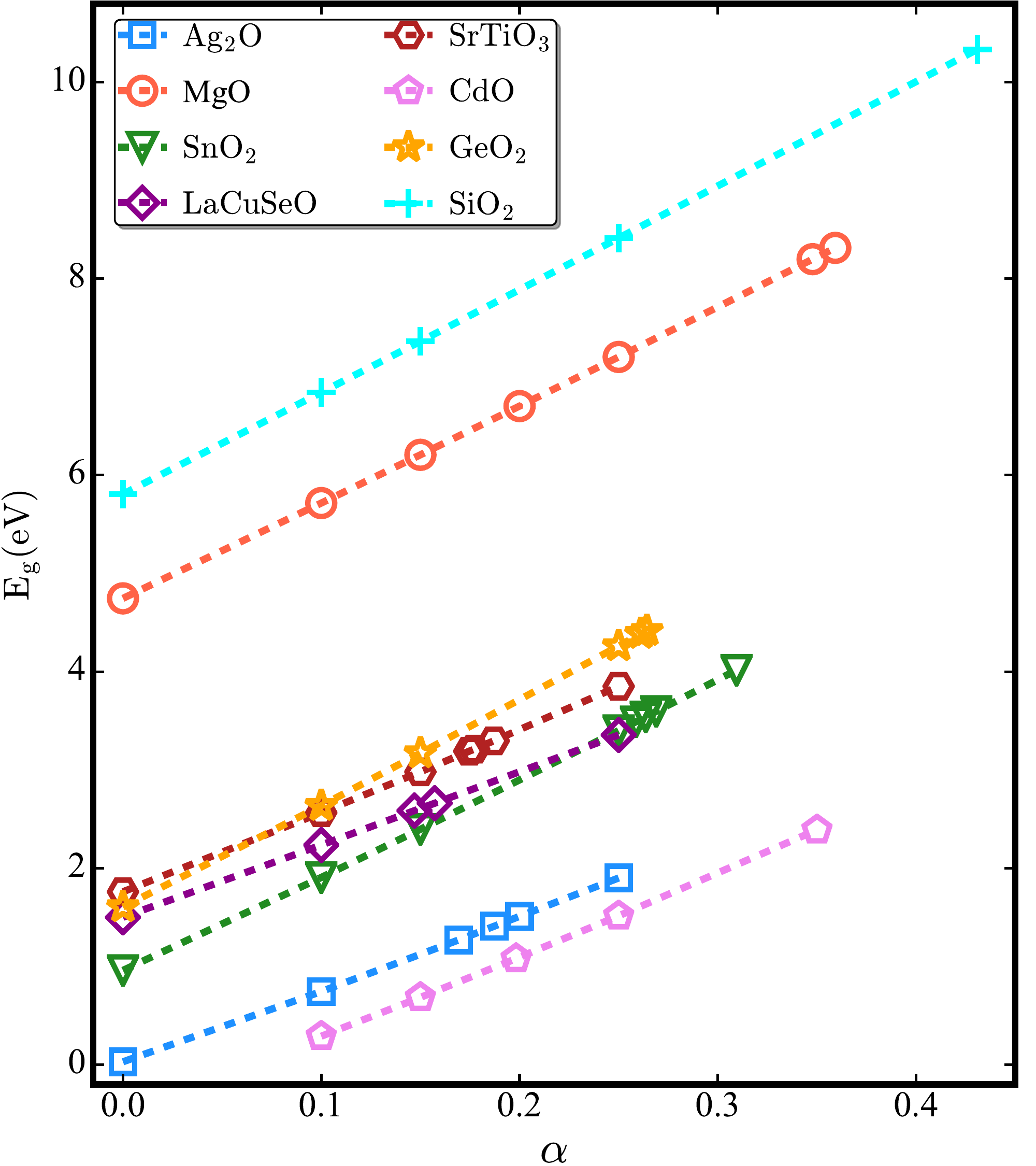}
        \caption
        {The dependence of band gap E$_g$  on the HF mixture parameter $\alpha$ for selected oxides.}
        \label{fig:01}
\end{figure}

Here the GGA functional is chosen according to the Perdew, Burke, Ernzerhof (PBE) parameterization~\cite{PBE}.
By analyzing the dependence of $E_{X,\lambda}$ on $\alpha$ and by a direct comparison with M{\o}ller-Plesset perturbation theory reference
energies, Perdew, Ernzerhof and Burke have found that the choice $\alpha$=0.25 yields the best atomization energies of typical
molecules~\cite{Perdew1996}. Although this 0.25 choice has become a standard in PBE0 calculations, the same authors have warned that
$\alpha$ is a system-specific and property-specific quantity. Indeed many researchers has treated $\alpha$ as an adjustable parameter to reproduce specific properties, in particular band gaps~\cite{PhysRevB.83.035119,alkauskas2011defect,PhysRevB.86.235117,PhysRevB.85.081109}.
It is known that the in hybrid functionals the size of the band gap scales linearly with 
$\alpha$~\cite{PhysRevB.86.235117,alkauskas2011defect}; we verify this scaling relation in Fig.~\ref{fig:01}(a) for selected semiconductors.

The PBE0 is the most representative example of a full-range hybrid, meaning that both the long-range (lr) and short-range (sr) part of the Coulomb potential are treated at HF level. The calculation of the slow-decaying lr part of the exchange integrals and exchange potential is numerically complicated and a large number of {\bf k} points is needed, leading to a very slow convergence. To solve this issue, Heyd, Scuseria, and Ernzerhof (HSE) have proposed to replace the lr-exchange by the corresponding density functional counterpart~\cite{Heyd2003a,Heyd2003b}, and to keep the HF description only in the short-range limit:

\begin{widetext}
\begin{equation}
\rm E_{XC}^{\rm HSE} = {\alpha}E_X^{HF,sr}(\mu) + (1-\alpha)E_X^{PBE,sr}(\mu) + E_X^{PBE,lr}(\mu) + E_C^{PBE}, 
\label{eq:hse}
\end{equation}
\end{widetext}

Here $\mu^{-1}$ is the critical screening length controlling the range separation, namely the distance at which the short-range Coulomb interactions can be assumed to be negligible.
Based on molecular tests the value of $\mu$ was set
to 0.2 \AA$^{-1}$ (corresponding to a screening length $r_s=2/\mu=10$ \AA), which is routinely
considered as the standard choice for range-separated HSE calculations~\cite{Heyd2003b}.

Recently there has been successful attempts to build parameter-free PBE0 hybrid, inspired by a comparison between the GKS  and the GW quasiparticle equations~\cite{Moussa2012,PSSB:PSSB201046303,DelSole1994,Gygy1989,PhysRevB.89.195112}. In the Hedin's GW equations, in fact, the XC functional is replaced by the self-energy $\Sigma$, evaluated in terms of the single particle Green function G and the screened Coulomb interaction W:

\begin{equation}
\Sigma(\omega) = i\rm{G}(\omega)\rm{W}(\omega),
\end{equation}

where and the screened Coulomb interaction

\begin{equation}
\rm{W}(\omega) = \epsilon(\omega)^{-1}v
\end{equation}

is expressed in terms of the bare Coulomb interaction $v=1/{|{\bf r}-{\bf r}'|}$ and the frequency dependent dielectric function $\epsilon$.
As already elaborated by other authors~\cite{Moussa2012,PhysRevB.89.195112,PhysRevB.83.035119}, in the GW static approximation
(known as the COHSEX approximation)~\cite{gw} one can replace the screening in $W$ by the static ($\omega=0$) macroscopic dielectric constant $\epsilon_{\infty}$, and, by setting $\epsilon_{\infty}=1/\alpha$, the COHSEX screened Coulomb potential W$^{COHSEX}$ become analogous to the corresponding  PBE0 screened Coulomb potential:

\begin{equation}
\rm{W}^{PBE0}(\omega=0)={\alpha}v.
\end{equation}
 
Using the relation $\alpha = 1/\epsilon_{\infty}$, some authors have recently proposed ``parameter-free`` full-range and range-separated hybrids~\cite{PhysRevB.83.035119,Shimazaki200891,doi:10.1063/1.3119259,PhysRevB.88.081204,0953-8984-25-43-435503,PhysRevB.89.195112,fritsch2017self,PhysRevB.93.235106, 10.1088/1361-648X/aa7823}. Some of these approaches are based on a fully self-consistent evaluation of $\epsilon_{\infty}$, which is computed at the end of each hybrid cycle using different theories and approximations~\cite{doi:10.1063/1.3119259,0953-8984-25-43-435503,PhysRevB.89.195112,fritsch2017self,10.1088/1361-648X/aa7823}. Among these, the protocol proposed by Skone \emph{et al.}, which adopts the coupled perturbed Kohn-Sham equations including local-field effects (i.e., density response due to external electric field) turned out to provide excellent agreement with experiments for the band gaps of a representative materials dataset composed by mono-atomic and binary compounds with dielectric constants spanning a wide range from 1.23 eV to 15.9 eV.\cite{PhysRevB.89.195112}  
The same authors have also proposed a range-separated version the same functional\cite{PhysRevB.93.235106}.

Inspired by the sc-hybrid of Skone and coworkers~\cite{PhysRevB.89.195112}, in this paper we employ a similar sc-hybrid and asses its performance for the prediction of band gaps in a wide class of binary, ternary and quaternary oxides, spanning band gaps between 0 to 12 eV, static dielectric functions within 3 and 15, and diverse structural properties. Our sc-hybrid is based on the 
Vienna \emph{Ab initio} Simulation Package (VASP)~\cite{gk1,gk2} and the dielectric function is computed from the response to finite electric fields 
using the perturbation expansion after discretization (PEAD) method~\cite{PhysRevB.63.155107,PhysRevLett.89.117602}. During the writing of our manuscript, we became aware of the work of Fritsch \emph{et al.} in which exactly the same type of hybrid is employed for calculations of the structural and electronic properties of three binary oxides: ZnO, MgO and SnO$_2$~\cite{fritsch2017self}.
Consistent with the nomenclature proposed by Fritsch \emph{et al.} we also call this functional scPBE0.
To assess the quality of the scPBE0 we have considered a wide material dataset formed by thirty different oxide semiconductors; 
our oxides dataset includes strongly correlated oxides~\cite{dagotto2008strongly}, transparent conducting oxides~\cite{zhang2016p}, thermoelectric~\cite{li2012high}, photocatalysis~\cite{linsebigler1995photocatalysis}, photovoltaic materials~\cite{sullivan2016copper}, ferroelectrics, refractory materials, and other oxides which are sued in the emerging field of oxide-based electronics. 

We will also compare the performance of scPBE0 with the standard screened HSE06 and full-range PBE0 results.
We would like to underline that the inverse dielectric constant relation for the estimation of the 'optimum' $\alpha$ is not directly applicable to range-separated hybrid functionals such as HSE06, because in screened hybrid functionals screening is already present to some extent as a consequence of the range separation~\cite{PhysRevB.86.235117}.
Considering that in HSE06 the Coulomb kernel is decomposed into a lr and sr part by way of the error function {\em erf}
and the complementary error function {\em erfc}~\cite{Heyd2003b, 0953-8984-26-25-253202}:

\begin{equation}
\frac{1}{r} =   \underbrace{ \frac{\mathrm {erfc}(\mu r)}{r} }_{\text{sr}}
+ \underbrace{  \frac{\mathrm {erf} (\mu r)}{r}}_{\text{lr}}
\end{equation}

the HSE06 screened Coulomb potential takes the form:

\begin{equation}
\rm{W}^{HSE(\omega=0)} = \alpha\frac{{erfc}{(\mu{|{\bf r}-{\bf r}'|})}}{|{\bf r}-{\bf r}'|}
\end{equation}
 
which clearly depends on the screening parameter $\mu$. In Ref.~\cite{PhysRevB.86.235117} the effect of the HSE06 screening 
on $\alpha$ is quantified as a downward shift of about 0.07 with respect to the 'optimum' $\alpha = 1/\epsilon_{\infty}$ value.

\section{Methodology and Computational Details}
The scPBE0 functional employed in this study is based on the full-range PBE0 functional as implemented in the VASP~\cite{gk1,gk2}, and the static dielectric function is evaluated using the perturbation expansion after discretization (PEAD) method~\cite{PhysRevB.63.155107,PhysRevLett.89.117602}, by considering the perturbation of a small but finite homogeneous electric field
$\upvarepsilon$ on the ground-state of an insulating system. In the VASP this is done by minimizing the total energy functional
$E[\{ \psi^{(\upvarepsilon)},\upvarepsilon \}]$ :

\begin{equation}
\rm{E}[\{ \psi^{(\upvarepsilon)},\upvarepsilon \}] = \rm{E_0}[\{ \psi^{(\upvarepsilon)}\}] - \Omega\upvarepsilon\cdot\textbf{P}[\{\psi^{(\upvarepsilon)}\}],
 \label{eq:electric}
\end{equation}

with respect to the field-polarized Bloch functions $\{ \psi^{(\upvarepsilon)}\}$;
in Eq.~\ref{eq:electric} $\Omega$ is the cell volume and $\textbf{P}[\{\psi^{(\upvarepsilon)}\}]$ is the macroscopic polarization as 
defined in the modern theory of polarization~\cite{PhysRevB.63.155107, PhysRevB.47.1651, RevModPhys.66.899}:

\begin{equation}
\textbf{P}[\{\psi^{(\upvarepsilon)}\}]=-\frac{2ie}{(2\pi)^3}\sum_n\int_{BZ}d\textbf{k} \langle u_{n\textbf{k}}^{(\upvarepsilon)} |\nabla_{\textbf{k}} | u_{n\textbf{k}}^{(\upvarepsilon)} \rangle,
\label{eq:mtp}
\end{equation}

where $u_{n\textbf{k}}^{(\upvarepsilon)}$ is the cell-periodic part of $\{ \psi_{n\textbf{k}}^{(\upvarepsilon)}\}$.
To compute the $\{ \psi^{(\upvarepsilon)}\}$ it is necessary to minimize the functional 
$E[\{ \psi^{(\upvarepsilon)},\upvarepsilon \}] = \langle \psi_{n\textbf{k}}^{(\upvarepsilon)} |H| \psi_{n\textbf{k}}^{(\upvarepsilon)} \rangle$,
which results in the following optimization problem:

\begin{equation}
 H| \psi_{n\textbf{k}}^{(\upvarepsilon)} \rangle =  
 H_0| \psi_{n\textbf{k}}^{(\upvarepsilon)} \rangle - \Omega\upvarepsilon\cdot 
 \frac{\delta \textbf{P}[\{\psi^{(\upvarepsilon)}\}]} {\delta\langle \psi_{n\textbf{k}}^{(\upvarepsilon)}|}
 \label{eq:min}
\end{equation}

which represents the gradient of the functional in Eq.~\ref{eq:electric}.
In the PEAD scheme the dielectric tensor is given by~\cite{PhysRevLett.89.117602}:

\begin{equation}
\epsilon_{ij} = \delta_{ij} + \chi_{ij} = \delta_{ij} + 4\pi\frac{\partial P_i}{\partial \upvarepsilon_j} 
 \label{eq:pead}
\end{equation}

where $\chi_{ij}=4\pi\frac{\partial P_i}{\partial \upvarepsilon_j}$ is the susceptibility. If the atoms are kept fixed, as in our case, 
this formula finally yields the so called \emph{ion-clamped} static dielectric tensor $(\epsilon_\infty)_{ij}$:

\begin{equation}
(\epsilon_\infty)_{ij} = \delta_{ij} + 4\pi\frac{(\textbf{P}[\{\psi^{(\upvarepsilon)}\}] - (\textbf{P}[\{\psi^{(0)}\}])_i}{\upvarepsilon_j}
 \label{eq:epsilon}
\end{equation}

In this formalism local-field effects are naturally taken into accounted through self-consistently.

\subsection{Technical Setup}
\begin{figure}
        \includegraphics[clip,width=0.99\linewidth]{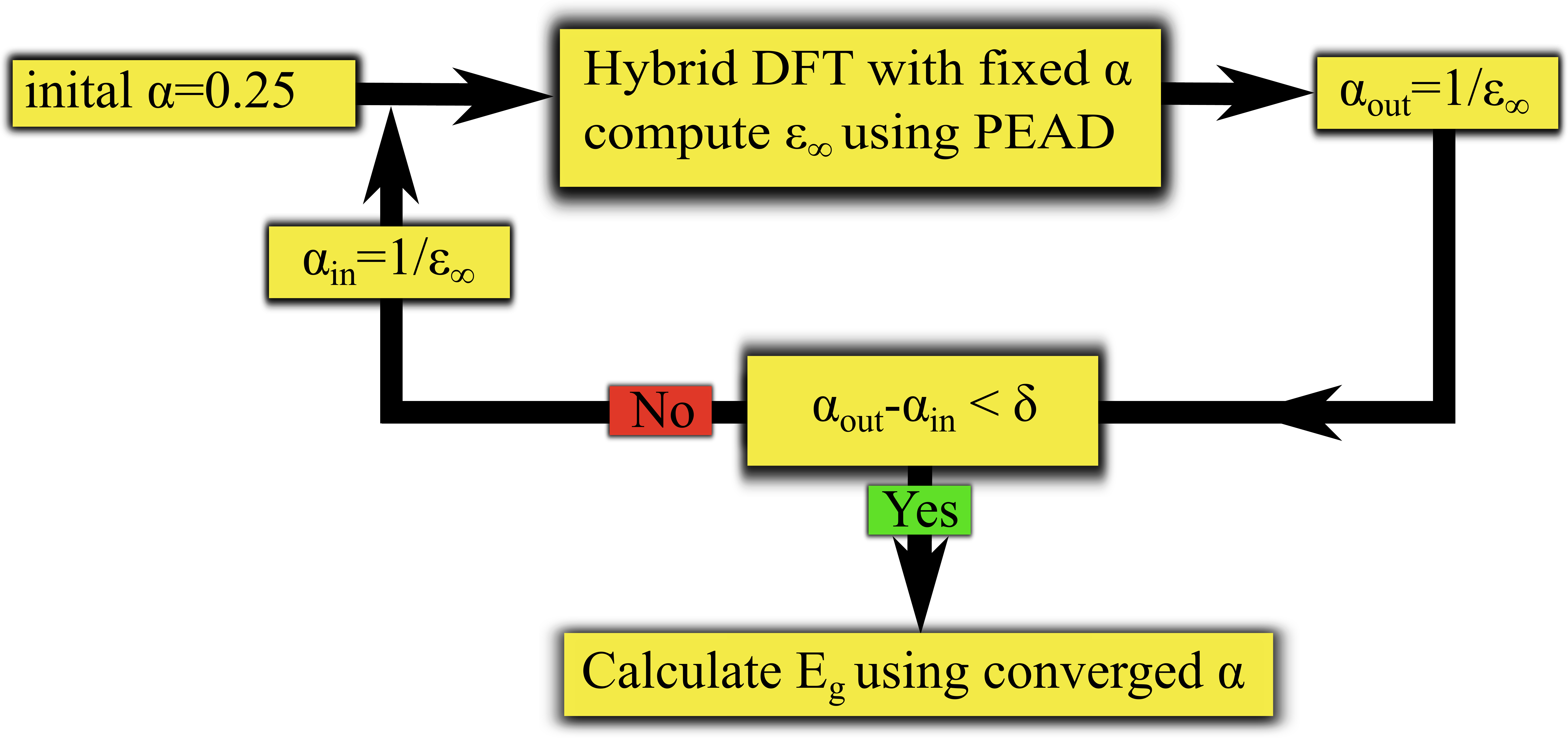}
        \caption
        {A flow chart of the self-consistent hybrid scheme.}
        \label{fig:02}
\end{figure}

All the calculations were performed using the projector augmented wave (PAW) pseudopotential method~\cite{PAW1,PAW2}. 
The crystal structures of the compounds under scrutiny were fully optimized (lattice constants and atom positions) within the PBEsol parametrization for the  exchange-correlation functional~\cite{PBEsol} adopting an energy cutoff for the plane wave expansion of 520 eV and a \emph{k}-mesh with KPPRA (k-point density per reciprocal atom) of 8000.
For all structures considered in our study we have used \emph{k}-meshes with KPRRA >3000 and energy cutoff of 400 eV for all hybrid functional calculations.
Based on the optimized structures, the dielectric function and the band gap have been computed at PBE0+PEAD level using the self-consistent protocol schematized in Fig.~\ref{fig:02}. 

\begin{figure}
	\includegraphics[clip,width=1.0\linewidth]{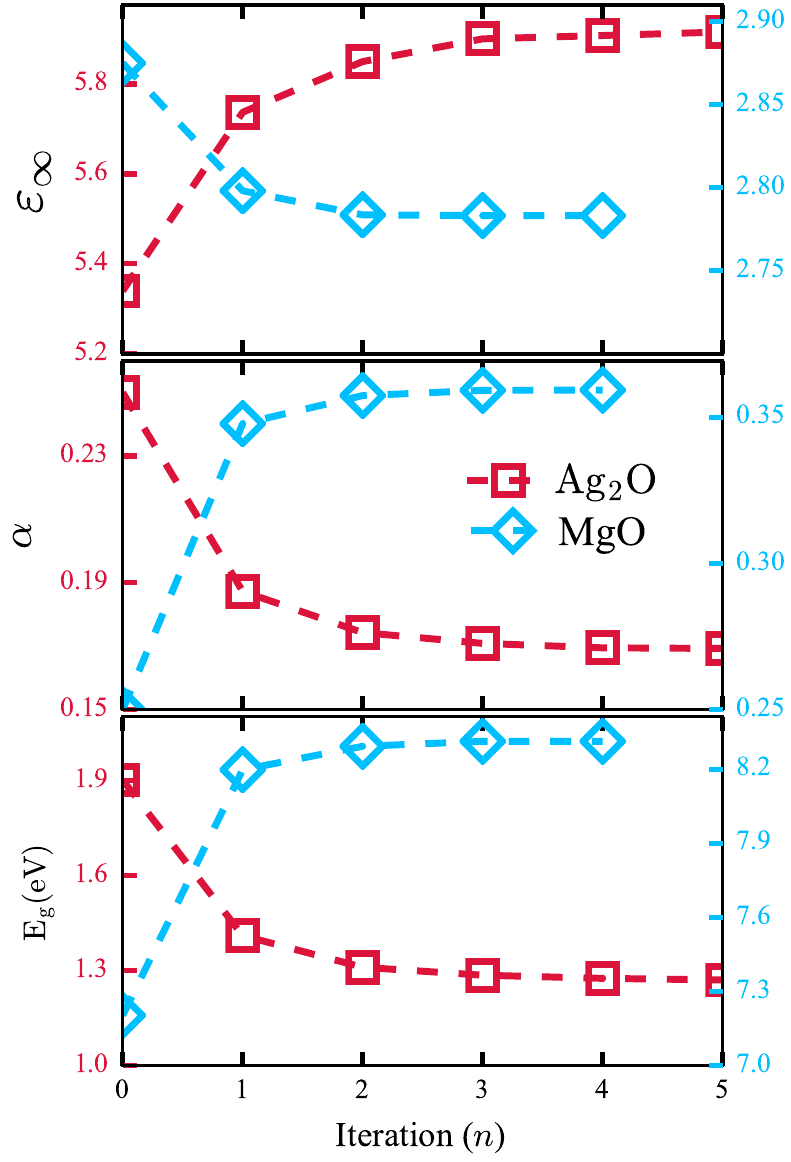}
	\caption
	{Evolution of the epsilon ($\overline{\varepsilon_{\infty}}$), alpha ($\alpha$), and band gap (${\rm{E}_g}$) of Ag$_2$O and MgO as a function of the number of iterations.}
	\label{fig:03}
\end{figure}

Specifically, we started the self-consistent loop by setting $\alpha_{\rm in}$=0.25, the standard mixing parameter for PBE0~\cite{adamo1999toward}. 
At the end of this iteration the new $\alpha$, $\alpha_{\rm out}$, has been determined as the inverse of the mean value of the diagonal elements of the dielectric function evaluated by PEAD, i.e., $\alpha_{\rm out}=1/\overline{\varepsilon_{\infty}}$.
This new value of $\alpha$ was then used to start a new iteration. We have continued this loop until  
$\alpha_{\rm out} - \alpha_{\rm in} < 0.01$. When the criterion is satisfied, the converged values of 
$\alpha$, $\epsilon_\infty$ as well as the band gap E$_{\rm g}$ are obtained.  
In Fig.~\ref{fig:03} we show the practical application of this self-consistent loop for the small band gap oxide Ag$_2$O and for the large band gap oxide MgO: the various quantities ($\alpha$, $\overline{\varepsilon_{\infty}}$, and E$_{\rm g}$) rapidly reach a well-converged value after only 4-5 iterations. Considering that the computation of the dielectric properties within the PEAD formalism is about 4 times more expensive than a standard PBE0 cycle, overall the scPBE0 procedure is about 20 times more expensive than standard non-self-consistent PBE0.

\begin{center}
\begin{table*}
	\caption{Band gaps (E$_{\rm g}$, in eV) and the averaged ion-clamped static dielectric tensor, $\overline{\varepsilon_{\infty}}$,
	for our oxides dataset calculated by scPBE0, PBE0, and HSE06. $\alpha_{\rm{sc}}$ and $\alpha_{\rm{fit}}$ are the
	self-consistent and fitted value of the mixing parameter $\alpha$. $\overline{\varepsilon_{\infty}}$ under scPBE0, fitted, HSE06, and PBE0
	are high-frequency dielectric constants calculated by using scPBE0, the fitted $\alpha$ with experiment ($\alpha_{\rm{fit}}$), HSE06, and PBE0, respectively.
    The experiment approaches used for band gap measurement are indicated under Method (TR: thermoreflectance; OA: optical absorption; XPS-BIS: x-ray photoelectron spectroscopy and bremsstrahlung isochromat spectroscopy; UPS: Ultraviolet photoelectron spectroscopy; SSD: Single scattering distribute spectra; PC: Photoconductivity; PE: photoemission; DRS: Diffuse reflectance spectra; ET: electrical transport).
	The last row show the mean absolute percentage errors (MAPE) for the calculated  band gaps and dielectric constants.
	For the evaluation of the bandgap-MAPE the data relative to Ag$_2$PdO$_2$  have been excluded due to the exceedingly large relative errors (statistically not relevant).  The materials are ordered by their chemical formula, from binary to quaternary compounds.
	} \vspace{0.3cm}
\begin{ruledtabular}
\begin{tabular}{cccccccccccccccc}
                    &               &    \multicolumn{3}{c}{scPBE0}  & \multicolumn{2}{c}{fitted}     &      \multicolumn{2}{c}{HSE06}             &        \multicolumn{2}{c}{PBE0}                   &             \multicolumn{2}{c}{Experiment}                                                                                                                                                                                                              \\
\cline{3-5} \cline{6-7} \cline{8-9} \cline{10-11} \cline{12-14}
   Compound         &Space group    &E$_{\rm{g}}$(eV)  &$\overline{\varepsilon_{\infty}}$&$\alpha_{\rm{sc}}$& $\alpha_{\rm{fit}}$ & $\overline{\varepsilon_{\infty}}$ & E$_{\rm{g}}$ (eV)  & $\overline{\varepsilon_{\infty}}$ &E$_{\rm{g}}$ (eV)& $\overline{\varepsilon_{\infty}}$& E$_{\rm{g}}$ (eV)                              & Method      & $\overline{\varepsilon_{\infty}}$&                                                                      \\
   CdO              & Fm$\bar{3}$m  &   0.99           &          5.35                   &    0.187         &         0.169       &         5.45                      &        0.88        &         5.12                      &  1.52        &        5.04                      &  0.84\cite{sringermaterial}                           & TR          &     6.2\cite{PhysRev.182.891}    &                                                                      \\
   PbO              & P4/nmm        &   2.13           &          5.86                   &    0.170         &         0.152       &         6.09                      &        1.93        &         5.55                      &  2.57        &        5.32                      &  2.03\cite{sringermaterial}                           & OA          &     7.1\cite{van1969optical}     &                                                                      \\
   MgO              & Fm$\bar{3}$m  &   8.31           &          2.78                   &    0.359         &         0.319       &         2.82                      &        6.48        &         2.91                      &  7.20        &        2.88                      &  7.9\cite{madelung2004ii}                             & OA          &     2.94\cite{madelung2004ii}    &                                                                      \\
   ZnO              & P6$_3$mc      &   3.38           &          3.63                   &    0.276         &         0.283       &         3.60                      &        2.49        &         3.74                      &  3.15        &        3.67                      &  3.44\cite{madelung2004ii}                            & OA          &     3.72\cite{madelung2004ii}    &                                                                      \\
   BeO              & P6$_3$mc      &  11.31           &          2.86                   &    0.349         &         0.282       &         2.91                      &        9.52        &         2.93                      &  10.25       &        2.92                      &  10.59\cite{sringermaterial}                          & OA          &     2.95\cite{sringermaterial}   &                                                                      \\ 
   NiO              & Fm$\bar{3}$m  &   3.51           &          6.60                   &    0.152         &         0.198       &         5.75                      &        4.45        &         5.13                      &  5.20        &        5.07                      &  4.30\cite{PhysRevLett.53.2339}                       & XPS-BIS     &     5.77\cite{pecharroman1994method}                                                                    \\
   SnO$_2$          & P42/mnm	    &   3.60           &          3.86                   &    0.269         &         0.269       &         3.86                      &        2.71        &         3.92                      &  3.40        &        3.23                      &  3.6\cite{sringermaterial}                            & UPS         &     4.06\cite{sringermaterial}   &                                                                      \\
   SiO$_2$          & P3$_1$21      &  10.49           &          2.24                   &    0.446         &         0.306       &         2.30                      &        7.67        &         2.32                      &  8.41        &        2.32                      &  9.0\cite{PhysRevB.83.174201}                         & OA          &     2.3\cite{malyi2016volume}    &                                                                      \\
   GeO$_2$          & P4$_2$/mnm    &   4.42           &          3.77                   &    0.264         &         0.345       &         3.63                      &        3.56        &         3.80                      &  4.26        &        3.83                      &  5.35\cite{madelung2004ii}                            & OA          &     4.43\cite{madelung2004ii}    &                                                                      \\
   HfO$_2$          & P4$_2$/nmc    &   6.83           &          4.46                   &    0.224         &         0.123       &         4.70                      &        6.34        &         4.39                      &  7.07        &        4.39                      &  5.9\cite{cheynet2007crystal}                         & SSD         &                                  &                                                                      \\
   TiO$_2$          & P4$_2$/mnm    &   2.90           &          7.07                   &    0.142         &         0.159       &         6.98                      &        3.12        &         6.45                      &  3.86        &        6.41                      &  3.05\cite{robertson2000band,PhysRev.87.876}          & ET          &     7.37\cite{sringermaterial}   &                                                                      \\              
   Ag$_2$O          & Pn$\bar{3}$m  &   1.27           &          5.92                   &    0.169         &         0.160       &         5.93                      &        1.21        &         5.70                      &  1.90        &        5.34                      &  1.20\cite{xu2000absolute,fortin1964photoconductivity}& PC          &                                  &                                                                      \\
   Cu$_2$O          & Pn$\bar{3}$m  &   1.77           &          7.22                   &    0.139         &         0.182       &         6.80                      &        2.07        &         6.34                      &  2.80        &        6.28                      &  2.17\cite{madelung2004ii}                            & OA          &     7.11\cite{madelung2004ii}    &                                                                      \\  
   Al$_2$O$_3$      & R$\bar{3}$c   &   9.63           &          2.98                   &    0.337         &         0.258       &         3.02                      &        7.99        &         3.03                      &  8.72        &        3.03                      &  8.8\cite{robertson2000band}                          & OA          &     3.4\cite{robertson2000band}  &                                                                      \\
   La$_2$O$_3$      & P$\bar{3}$m1  &   6.13           &          4.03                   &    0.249         &         0.216       &         4.10                      &        3.92        &         4.05                      &  6.14        &        4.02                      &  5.8\cite{shang2004stability}                         & OA          &                                  &                                                                      \\
   In$_2$O$_3$      & R$\bar{3}$c   &   3.32           &          4.10                   &    0.244         &         0.214       &         4.01                      &        2.72        &         3.68                      &  3.38        &        3.92                      &  3.02\cite{PhysRevB.79.205211}                        & OA          &     3.62\cite{prathap2006optical}&                                                                      \\   
   CuAlO$_2$        & R$\bar{3}$m   &   4.01           &          4.45                   &    0.226         &         0.120       &         4.82                      &        3.49        &         4.42                      &  4.24        &        4.37                      &  2.99\cite{pellicer2006band}                          & OA          &                                  &                                                                      \\                                                    
   LiCoO$_2$        & R$\bar{3}$m   &   4.22           &          4.79                   &    0.209         &         0.116       &         5.47                      &        4.12        &         4.64                      &  4.89        &        4.63                      &  2.7\cite{PhysRevB.44.6090}                           & XPS-BIS     &                                  &                                                                      \\
   LaAlO$_3$        & R$\bar{3}$c   &   6.27           &          4.08                   &    0.246         &         0.254       &         4.16                      &        5.57        &         4.05                      &  6.30        &        4.06                      &  6.33\cite{cicerrella2005optical}                     & OA          &     4.0\cite{nunley2016optical}  &                                                                      \\                          
   LiNbO$_3$        & R3c           &   5.46           &          4.48                   &    0.223         &         0.011       &         5.17                      &        4.97        &         4.41                      &  5.71        &        4.32                      &  3.50\cite{xu2000absolute}                            & OA          &     4.87\cite{PhysRev.158.433}   &                                                                      \\
   BiFeO$_3$        & R3c           &   2.87           &          7.03                   &    0.142         &         0.125       &         7.17                      &        3.40        &         6.12                      &  4.14        &        6.10                      &  2.67\cite{PhysRevB.79.134425}                        & OA          &     5.52\cite{li2007infrared}    &                                                                      \\
   BaTiO$_3$        & P4mm          &   3.11           &          5.54                   &    0.181         &         0.197       &         5.88                      &        3.05        &         5.54                      &  3.77        &        5.52                      &  3.26\cite{sringermaterial}                           & OA          &     5.75\cite{huang2006temperature}&                                                                    \\
   PbTiO$_3$        & P4mm          &   2.51           &          7.31                   &    0.138         &         0.261       &         6.29                      &        2.62        &         6.58                      &  3.32        &        6.43                      &  3.4\cite{robertson2000band,pandey2005structural}     & OA          &     6.25\cite{robertson2000band} &                                                                      \\
   BaSnO$_3$        & Pm$\bar{3}$m  &   3.15           &          3.98                   &    0.251         &         0.246       &         3.34                      &        2.45        &         4.25                      &  3.14        &        3.98                      &  3.1\cite{mizoguchi2004probing}                       & OA          &     3.3\cite{stanislavchuk2012electronic}&                                                              \\
   SrTiO$_3$        & Pm$\bar{3}$m  &   3.19           &          5.75                   &    0.175         &         0.188       &         5.59                      &        3.33        &         5.35                      &  3.85        &        5.52                      &  3.3\cite{robertson2000band,van2001bulk}              & OA          &     6.1\cite{robertson2000band}  &                                                                      \\
   LaMnO$_3$        & Pmna          &   2.12           &          5.67                   &    0.176         &         0.141       &         6.14                      &        2.27        &         5.16                      &  3.01        &        5.11                      &  1.7\cite{PhysRevB.51.13942}                          & PE          &     4.9\cite{arima1995optical}   &                                                                      \\  
   BiVO$_4$         & C2/c          &   2.95           &          6.68                   &    0.149         &         0.072       &         7.16                      &        2.97        &         6.41                      &  3.67        &        6.34                      &  2.4\cite{dunkle2009bivo4}                            & DRS         &                                  &                                                                      \\
   Ag$_2$PdO$_2$    & Immm          &   0.86           &          8.49                   &    0.115         &         0.026       &         11.05                     &        1.19        &         7.39                      &  1.89        &        7.34                      &  0.18\cite{schreyer2001synthesis}                     & ET          &                                  &                                                                      \\
   BiCuSeO          & P4/nmm        &   0.88           &         12.56                   &    0.080         &         0.067       &         12.25                     &        1.31        &        10.46                      &  1.92        &       10.18                      &  0.8\cite{hiramatsu2007crystal}                       & OA          &                                  &                                                                      \\
   LaCuSeO          & P4/nmm        &   2.59           &          6.85                   &    0.147         &         0.157       &         6.80                      &        2.68        &         6.43                      &  3.36        &        6.36                      &  2.8\cite{hiramatsu2007crystal}                       & OA          &                                  &                                                                      \\
   MAPE  (\%)       &               &   14.3           &          10.0                   &                  &                     &                                   &        18.5        &          8.8                      &  31.1        &        9.7                       &                                                       &                            &                                  &                                                                      \\
\end{tabular}
\flushleft
\end{ruledtabular}
	\label{crystal}
\end{table*}
\end{center}


\section{Results and discussion} 
\begin{figure}[h]
	\includegraphics[clip,width=1.0\linewidth]{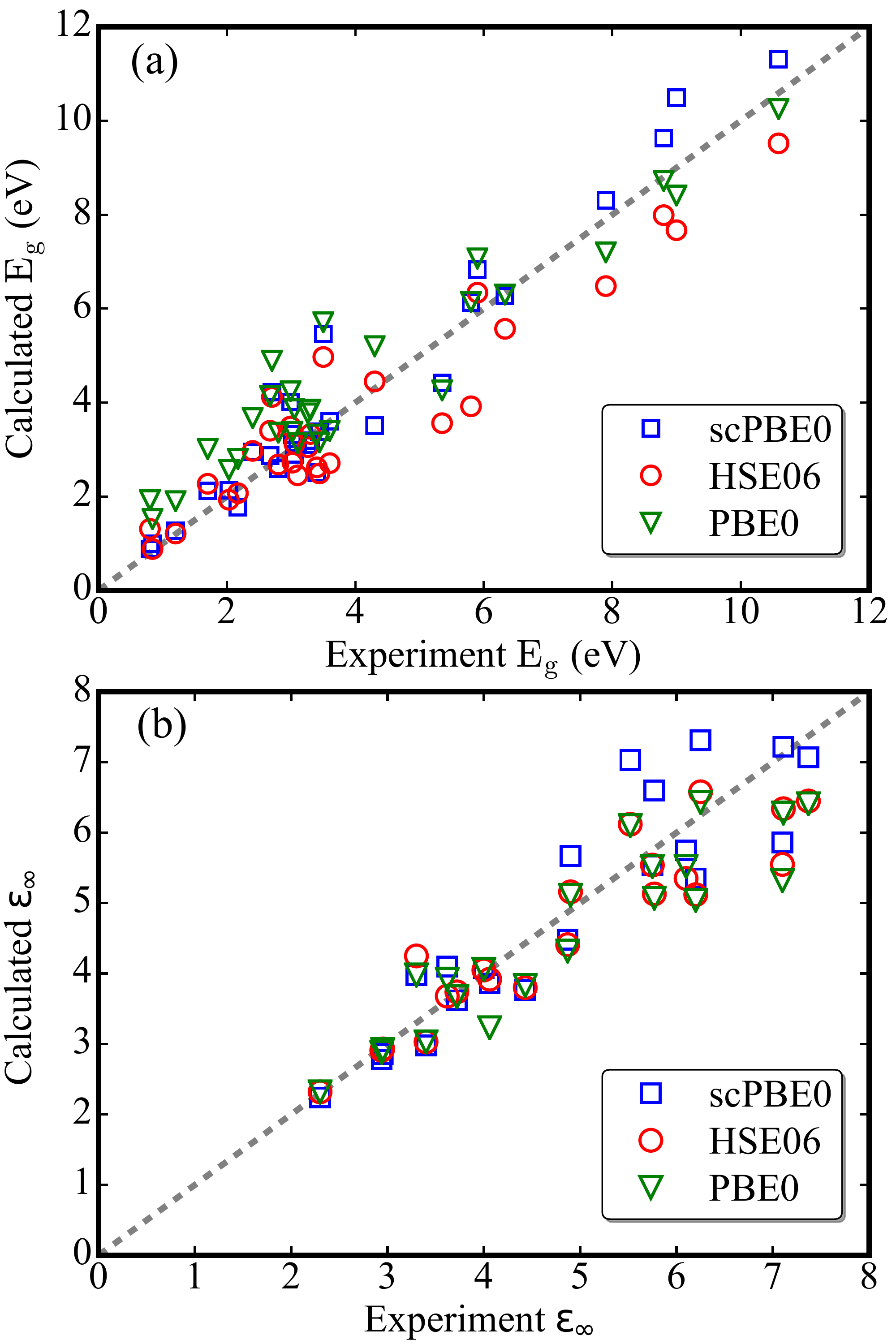}
	\caption
	{Comparison between calculated (scPBE0, PBE0, and HSE06) and the experiment band gaps (a) and ion-clamped dielectric constants (b).}
	\label{fig:04}
\end{figure}

To systematically evaluate the performance of the PEAD-based scPBE0 functional we have computed the bandgap and ion-clamped  dielectric constant of thirty different oxide semiconductors using scPBE0, PBE0 and HSE06, and compared the results with available experimental data. The data are collected in Tab.~\ref{crystal}, where we also provide the list of all compounds scrutinized in our study; a graphical summary of the calculated data is provided in Fig.~\ref{fig:04}.

\begin{figure*}
	\includegraphics[clip,width=1.0\linewidth]{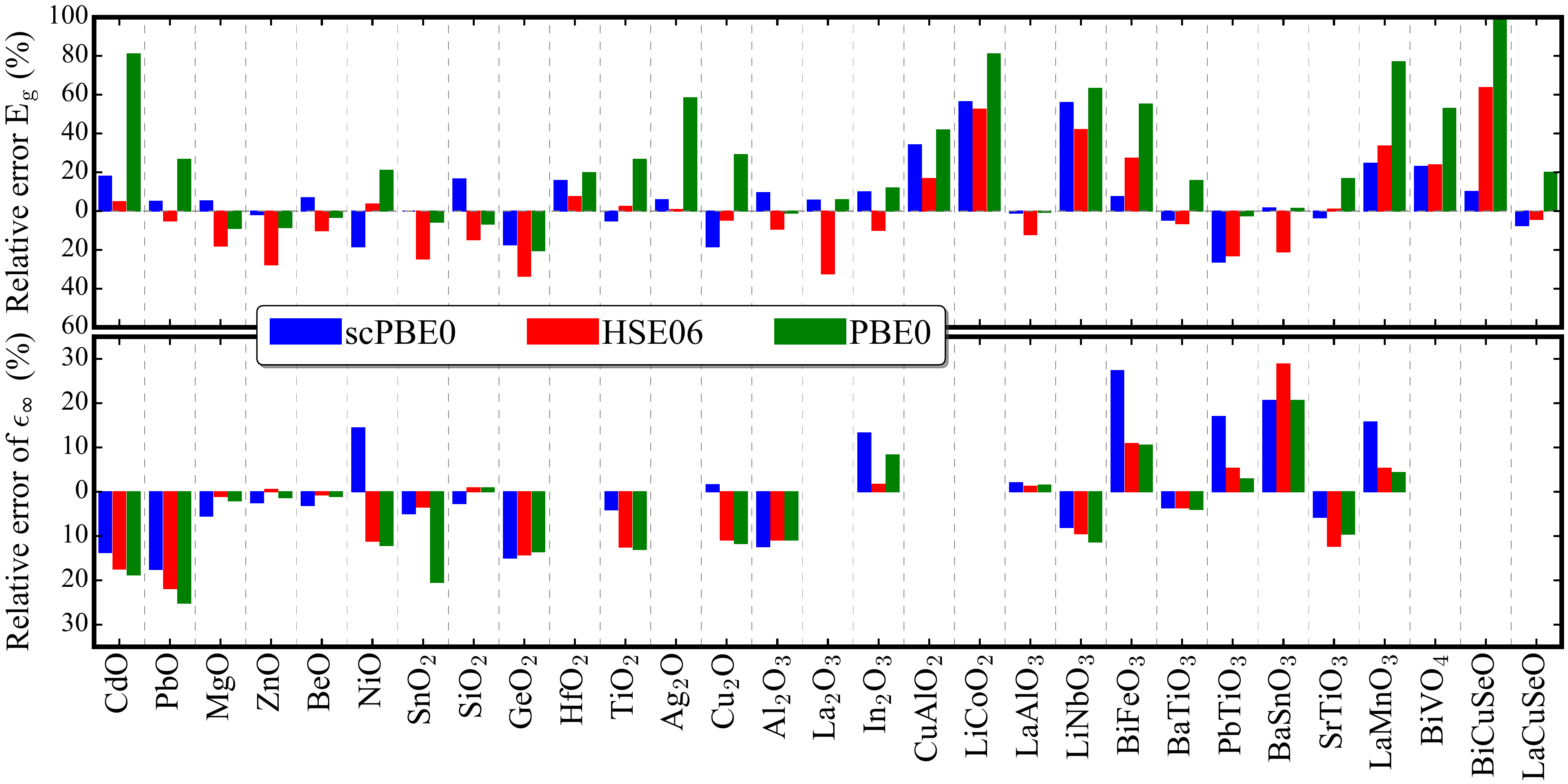}
	\caption
	{The relative errors of the calculated (scPBE0, HSE06, and PBE0) band gap (upper panel) and dielectric constant (lower panel) 
	with respect to the experimental values.}
	\label{fig:05}
\end{figure*}

To quantify the relative predictive power of the employed computational schemes we have also computed the relative errors for all considered semiconductors, which are shown as histograms in 
Fig.~\ref{fig:05}, as well as the mean absolute percentage error defined by the formula:
\begin{equation}
\rm{M^{theory}_{E_g}} = \frac{1}{N} \sum_{i=1}^N \left|\frac{\rm E_g^{expt}-E_g^{theory}}{\rm E_g^{expt}}  \right| \times 100\%
\label{eq:mare}
\end{equation}
where N=30 is the total number of compounds considered, and \emph{theory} refers to the specific type of functional used in the calculation. A similar formula has been used for evaluating the MAPE relative to $\overline{\varepsilon_{\infty}}$ ($\rm{M^{theory}_{\overline{\varepsilon_{\infty}}}}$).
The value of the MAPE, collected 
in the last row of Tab.~\ref{crystal} indicate that scPBE0 is the most accurate method for the evaluation of the band gaps, with an overall MAPE 
M$^{\rm {scPBE0}}_{\rm E_g}$=14.3 \%, which is smaller than the corresponding HSE06 (M$^{\rm {HSE06}}_{\rm E_g}$= 18.5\%), and PBE0 (M$^{\rm PBE0}_{\rm E_g}$= 31.1\%) values. On the other side, for what concern the ion-clamped  dielectric function, the three methods are almost equivalent with PBE0 delivering slightly (1\%) better estimations:
M$^{\rm scPBE0}_{\overline{\varepsilon_{\infty}}}$=10 \%, M$^{\rm PBE0}_{\overline{\varepsilon_{\infty}}}$= 8.8\%), and M$^{\rm HSE06}_{\overline{\varepsilon_{\infty}}}$= 9.7\%.

The calculated data indicate that scPBE0 performs better than HSE06 and PBE0 for band gaps smaller than $\approx$ 8 eV, whereas for larger gaps PBE0 appears to be a better choice. The best performance of scPBE0 with respect to the fixed-$\alpha$ PBE0 functional is exemplified in Fig.~\ref{fig:03}:
for small band gap materials such as Ag$_2$O (1.20 eV) $\alpha$ decreases monotonically with increasing number of iteration and finally reaches a converged value of 0.169, significantly smaller than the standard PBE0 value, 0.25; in the larger band gap limit, represented by MgO (E$_{\rm g}$ = 7.9 eV), the self-consistent value of $\alpha$, 0.359, is, on the other hand, larger than 0.25. 

Apparently, there are exceptions to this general trend: for 
BeO (E$_{\rm g}$ = 10.59 eV), SiO$_2$ (E$_{\rm g}$ = 9.0 eV),  and Al$_2$O$_3$ ($E_g$ = 8.8 eV) scPBE0 predicts larger band gaps than PBE0 
due to the larger value of the self-consistent $\alpha$ as compared to the standard 0.25 choice. However, for these low dielectric constant materials ($\approx$ 3) excitonic effects, not taken into account in the present study, can be very strong and this unavoidably
affect the comparison between calculated and measured data. For example in SiO$_2$ the excitonic binding energy is as larger as 1.2 eV~\cite{PhysRevB.85.045205} and the measured band gap is 9.0 eV; by taking into account this excitonic shift in our calculations, scPBE0 turns out to provide the best prediction of the band gap, 9.29 eV, whereas PBE0 and HSE are off by more than 2 eV (E$_{\rm g}^{\rm PBE0}$ = 7.21 eV and E$_{\rm g}^{\rm HSE06}$ = 6.47 eV). Therefore, the fact that scPBE0 gives band gaps larger than experiments for materials with a weak dielectric screening 
(i.e., typically large excitonic effects) should be considered as a positive aspect. The inclusion of excitonic effects would lead to a decrease of the band gap thus improving the comparison with experiment. Unfortunately, measuring or calculating  excitonic binding energies is an extremely complicated task~\cite{PhysRevLett.99.246403}  and only very few values are available in literature for the compounds included in our materials dataset. 
The materials for which scPBE0 overestimates the band gaps with respect to the measured values are: BeO, SiO$_2$, HfO$_2$, Al$_2$O$_3$, CuAlO$_2$, LiCoO$_2$, NaSbO$_3$, and LiNbO$_3$. Among these materials, to our knowledge, the excitonic binding energy is accurately reported only for 
SiO$_2$~\cite{PhysRevLett.99.246403}.

In the small band gap limit (usually large $\epsilon_{\infty}$) excitonic effects are expected to be week (from few tens to few hundreds of meV), 
scPBE0 performs much better than PEB0. It is worth noting that for small band gaps in the range 0 $\sim$ 3 eV, the screened HSE06 functionals performs relatively well, due to the intrinsic screening incorporated in this type of functionals via the range-separation which allow for a better description of the metallic limit. Considering that no extra screening is adopted in scPBE0, a comparable performance with HSE06 for these compounds indicates a success of scPBE0 functional for small band gap oxides.

Lastly, we would like to mention that we found that scPBE0 breaks down for two oxides with small band gaps, i.e., PdO (E$_{\rm{g}}$=1.0 eV~\cite{xu2000absolute}) and AgBiO$_3$ (E$_{\rm{g}}$=0.8 eV~\cite{mizoguchi2004electronic}): the self-consistent procedure in these two cases end up with a metallic solution due to the a considerable overestimation of $\epsilon_{\infty}$.

\begin{figure}
	\includegraphics[clip,width=1.0\linewidth]{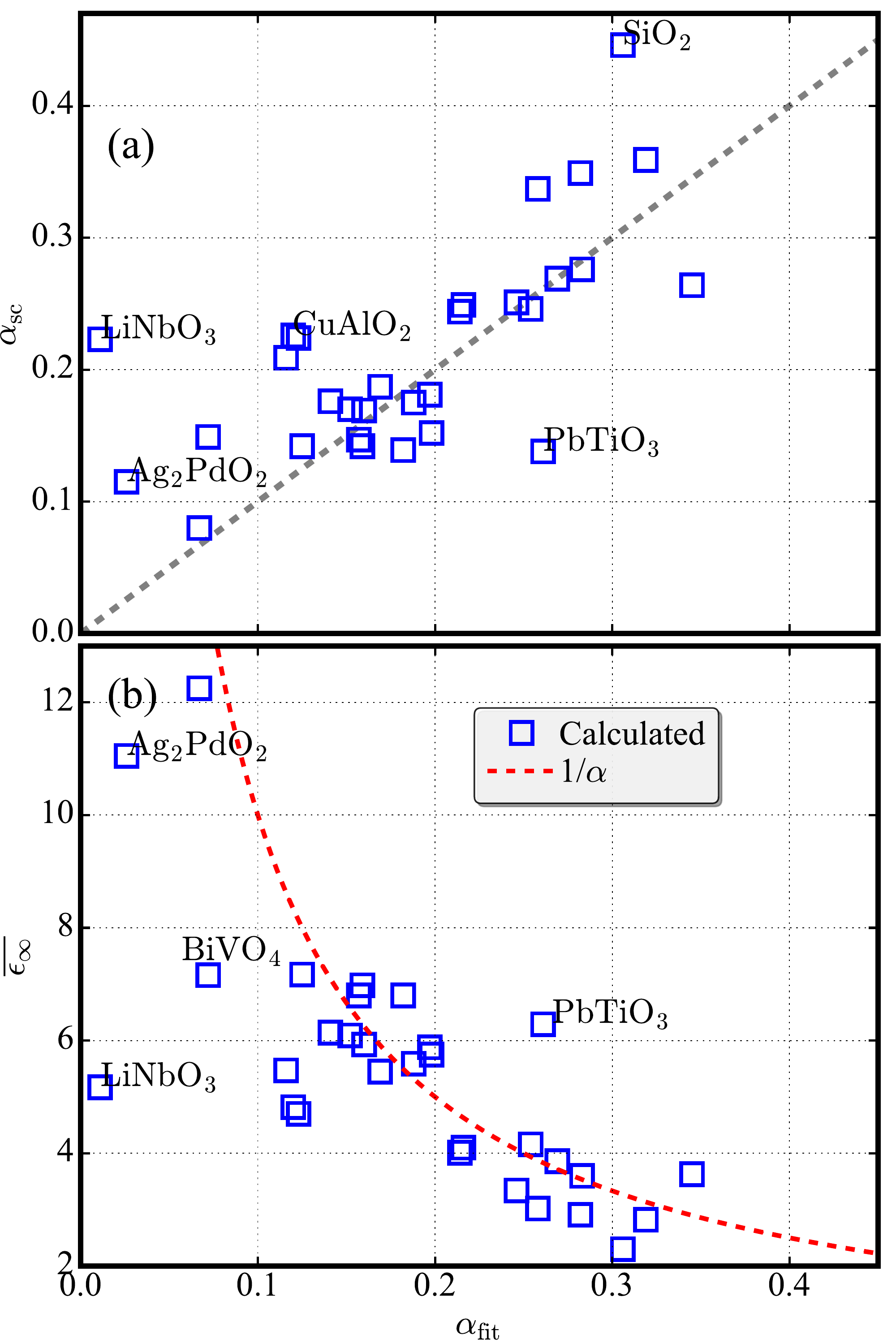}
	\caption
	{(a) Comparison between the self consistent value of $\alpha$ obtained by scPBE0 ($\alpha_{\rm{sc}}$, symbols) and the fitted 
	values obtained by the linear relation shown in Fig.\ref{fig:01} ($\alpha_{\rm{fit}}$, dashed line)); (b)
	Computed values of $\overline{\varepsilon_{\infty}}$ obtained at PBE0-PEAD level using the self-consistent value of $\alpha$,
	$\alpha_{sc}$. 
	The dashed-line indicates $\overline{\varepsilon_{\infty}}$ according to the (non-linear) relation $\alpha=1/\overline{\varepsilon_{\infty}}$ used for the self-consistent procedure.
	}
	\label{fig:06}
\end{figure}

As mentioned in the introduction, a practical remedy to improve the agreement with experiment within a non-self-consistent hybrid framework 
is to treat $\alpha$ as a fitting parameter,
for instance by setting $\alpha$ to the value that reproduce the measured band gap, $\alpha_{\rm fit}$. 
This can be done by using the linear relation between E$_{\rm{g}}$ and $\alpha$ depicted in Fig.~\ref{fig:01}. The resulting values of $\alpha_{\rm fit}$, also reported in Tab.~\ref{crystal}, ranges from 0.011 (LiNbO$_3$) to 0.345 (GeO$_2$); this wide range of variation of $\alpha_{\rm fit}$ is a further demonstration that this quantity is indeed material-dependent.
A direct comparison between $\alpha_{fit}$ and the corresponding self-consistent value adopted in scPBE0, $\alpha_{\rm sc}$, shown in Fig.~\ref{fig:06}(a), show strong deviations, with an overall tendency of scPBE0 to yield larger values of $\alpha$. This difference between scPBE0 and fitted-PBE0 is 
due to the fact that  $\alpha$ and ${\varepsilon_{\infty}}$ does not scale linearly~\cite{PSSB:PSSB201046195,0370-1301-63-3-302,moss1985relations}.
This is graphically shown in Fig.~\ref{fig:06}(b), where we report the values of  $\overline{\varepsilon_{\infty}}$ 
according to the relation $\alpha=1/\overline{\varepsilon_{\infty}}$ used for the self-consistent procedure,
as well as the values of $\overline{\varepsilon_{\infty}}$ obtained at PBE0-PEAD level using the self-consistent 
value of $\alpha$, $\alpha_{sc}$. 
As a result, by fitting $\alpha$ with respect to the band gap the MAPE associated with the band gaps is minimized but the description of the dielectric 
properties get worse. On the other hand we emphasize once more that scPBE0 is based on a self-consistent evaluation of the dielectric constant, which
guarantees a good description of the dielectric screening and,  at the same time, the self-consistent $\alpha$ obtained through the relation 
$\alpha = 1/\epsilon_{\infty}$ allows for a generally good prediction of the band gap.

The non-linearity between E$_{\rm g}$ and $\epsilon_{\infty}$ has been pointed out first by Moss~\cite{0370-1301-63-3-302,moss1985relations}, who proposed the formula, known as Moss relation:

\begin{equation}
 \epsilon_{\infty}^2 = \rm C \cdot {E_g}
\end{equation}

which has been verified for many elements and semiconducting compounds, including oxides\cite{REDDY1995825}, 
though deviations with respect to the measured values have been reported for band gap smaller than 1.4 eV~\cite{doi:10.1063/1.359248}.

\begin{figure}
	\includegraphics[clip,width=1.0\linewidth]{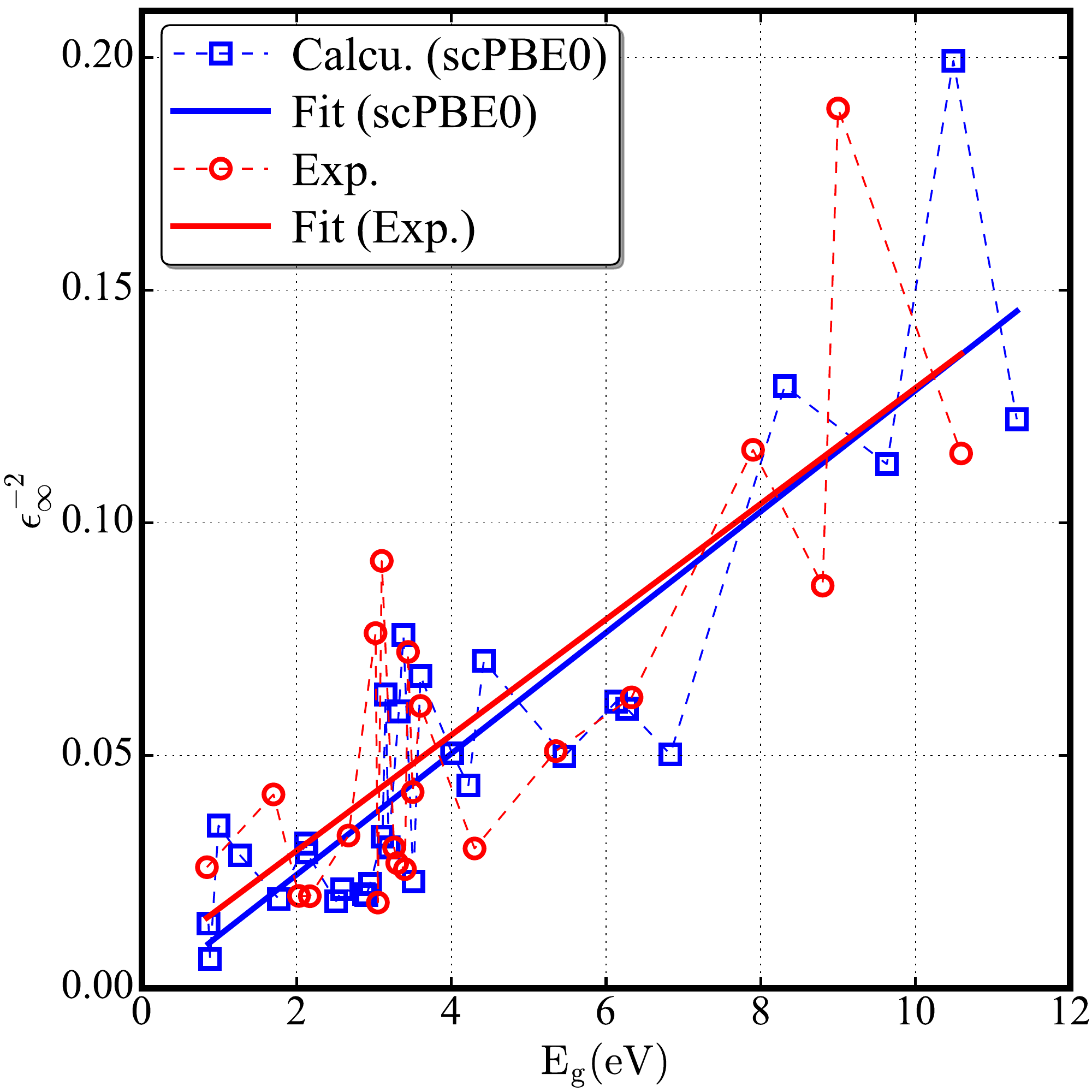}
	\caption
	{The relation between band gap ($E_g$) and dielectric constant ($\epsilon_{\infty}$) based on the scPBE0 and available experimental results.
	The solid lines are a linear fit (Moss relation).}
	\label{fig:07}
\end{figure}

In Fig.~\ref{fig:07} we show that the Moss relation is sufficiently well fulfilled for our material dataset,
not only experimentally but also based on the scPBE0 results: a linear fit of the measured and scPBE0 data leads to two almost overlapping 
lines with C$\approx$0.012.

\section{Summary and Conclusions}

In summary, in this paper we have assessed the performance of the unscreened self-consistent hybrid functional scPBE0 for a wide set of oxide materials with different dielectric, optical and structural characteristics.  
In the scPBE0 scheme the HF mixing parameter $\alpha$ is determined by a self-consistent evaluation of ion-clamped  dielectric constants using the PEAD method and by making use of the relation $\alpha = 1/\epsilon_{\infty}$. Therefore this method is \emph{de facto} an \emph{ab initio} approach, without adjustable or empirical parameters. We have compared the scPBE0 results with those obtained using the standard PBE0 functional (with a fix value of $\alpha$=0.25) and the screened (range-separated) HSE06 functional.
Our results shows that scPBE0 outperforms PBE0 and, to a lesser extent, HSE06 for the prediction of band gaps, with an overall MAPE of 14.3\%.
As for the prediction of the ion-clamped  dielectric constant all methods deliver similar results, associated wit ha MAPE of about 10 \%.

Importantly, for materials characterized by a weak dielectric screening and therefore prone to exhibit large excitonic effects, scPBE0 (unlike PBE0 and HSE06) furnishes band gaps larger than the measured data. The inclusion of excitonic effects (neglected in this study) would therefore  improve the agreement with experiment only for scPBE0; this has been verified for SiO$_2$, for which the excitonic binding energy is as large as 1.2 eV.  
On the other side, in the very small band gap limit, we found that scPBE0 could lead to metallic solutions. This was the case for PdO and AgBiO$_3$, which are better described at PBE0 and HSE06 level.

Moreover, we have shown that by taking advantage of the linear relation between $\alpha$ and E$_{\rm g}$ it is possible to obtain a set of optimally fitted $\alpha$ that guarantee an excellent agreement with experiment at PBE0 level. However, due to the non-linearity between $E_g$ and $\epsilon_{\infty}$,
with these fitted values of $\alpha$ the dielectric properties are not well described.
Finally, we have verified that the Moss relation ($\epsilon_{\infty}^2 = \rm C \cdot E_g$) is fulfilled for both experimental and scPBE0 set of data.

In conclusion, scPBE0 represents an valuable scheme to obtain a satisfactory description of band gaps and dielectric properties in materials fully \emph{ab initio}, thus representing a step forward with respect to parameter-dependent hybrid functionals. Clearly, to achieve a more accurate account of the optical and screening properties of materials (with error smaller than 2-3\%) it is necessary to go beyond the Hartree-Fock picture. A natural option is the GW approach with the inclusions of excitonic effects, which is however computationally much more demanding than hybrid functional schemes.

\section{ACKNOWLEDGMENTS}
This work was sponsored by the FWF project INDOX (Grant No.\ I1490-N19). All
calculations were performed on the Vienna Scientific Cluster (VSC).

\bibliography{Reference}

\end{document}